\documentclass[letter,letter]{jpsj2} 
\makeatletter
\newcommand{\contraction}[5][1ex]{%
  \mathchoice
    {\contraction@\displaystyle{#2}{#3}{#4}{#5}{#1}}%
    {\contraction@\textstyle{#2}{#3}{#4}{#5}{#1}}%
    {\contraction@\scriptstyle{#2}{#3}{#4}{#5}{#1}}%
    {\contraction@\scriptscriptstyle{#2}{#3}{#4}{#5}{#1}}}%
\newcommand{\contraction@}[6]{%
  \setbox0=\hbox{$#1#2$}%
  \setbox2=\hbox{$#1#3$}%
  \setbox4=\hbox{$#1#4$}%
  \setbox6=\hbox{$#1#5$}%
  \dimen0=\wd2%
  \advance\dimen0 by \wd6%
  \divide\dimen0 by 2%
  \advance\dimen0 by \wd4%
  \vbox{%
    \hbox to 0pt{%
      \kern \wd0%
      \kern 0.5\wd2%
      \contraction@@{\dimen0}{#6}%
      \hss}%
    \vskip 0.2ex
    \vskip\ht2}}

\newcommand{\contraction@@}[3][0.05em]{%
  \hbox{%
    \vrule width #1 height 0pt depth #3%
    \vrule width #2 height 0pt depth #1%
    \vrule width #1 height 0pt depth #3%
    \relax}}
\makeatother

\title{A macroscopic persistent current generation by merons in a spin density wave ordered state}

\author{Hiroyasu \textsc{Koizumi}}

\inst{Institute of Materials Science,University of Tsukuba,Tsukuba, Ibaraki 305-8573, Japan}

\abst{We show that a stable spontaneous current appears in a system of a spin density wave order with {\em merons}, vortices in the spin configuration with winding number $+1$. A meron in the spin density wave order modifies the boundary condition for eigenfunctions around it to a sign-change one. As a consequence, two types of stable current states, one with a clockwise circulation and the other with a counterclockwise one, arise. When a magnetic field is present, one produces a diamagnetic current is chosen. A collection of such currents results in a large diamagnetic current; and if the meron density is sufficiently large, a perfect diamagnetism is realized. The stability of this diamagnetic current is attributed to the topological nature of the merons, and as long as the distribution of the merons remains the same the  current will persist even in a macroscopic system.}

\kword{Berry phase, SDW, meron, persistent current}

\begin{document}

\maketitle
It is known that a quantum number for a circular motion is modified from integers to half-odd integers when the boundary condition on eigenfunctions is altered from a single-valued one to sign-change one by a Berry phase.\cite{Berry84}  Such a change was first found in the context of the dynamical Jahn-Teller effect,\cite{Longuet-Higgins58,Ham87} but now similar effects have been found in many other systems,\cite{Bohm03} including band twisting ones.\cite{Koizumi98,Koizumi99b} One of the important consequences brought about by this quantum number modification is that the ground state becomes degenerate.\cite{Ham87,Koizumi94,Koizumi99,Koizumi00}

\begin{figure}
\begin{center}
\includegraphics[scale=0.5]{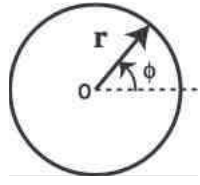}
\end{center}
\caption{\label{fig:ring} A ring-shaped system and the definition of $\phi$.}
\end{figure}

To see this, let us consider a particle in a ring-shaped system given in Fig.~\ref{fig:ring}. We denote the polar angle for the position in the ring by $\phi$. 
When the ordinary single-valued boundary condition is imposed, eigenfunctions for the particle state are
 $(2\pi)^{-1/2}e^{in\phi}$ ($n$ is an integer). The ground state is the $n=0$ state, which is not degenerate.  On the other hand, eigenfunctions for the sign-change boundary condition have the same functional form but with half odd integers for $n$; thus, the ground states are those with $n=\pm{1 \over 2}$ and degenerate.
If the particle has an electric charge, this degeneracy implies an electric current generation in the ground state. 
  
In the present work, we deal with a spin density wave (SDW) ordered state with {\em merons}; and examine the appearance of circular currents around them due to the sign-change boundary condition, a mechanism similar to the one explained above. 

Merons are vortices in the spin configuration with winding number $+1$ (if the winding number is $-1$ it is called, ``antimerons").\cite{John04}  A typical meron in the $xy$ plane is given by
\begin{eqnarray}
\eta({\bf r})=\tan^{-1}{{y-y_c} \over {x-x_c}},
\end{eqnarray}
where $(x_c, y_c)$ is the center of the meron, and $\eta$ is the angle for spin direction in the $xy$ plane.

There is compelling evidence that merons exist in the background antiferromagnetic spin order in the pseudogap phase of cuprate superconductors.\cite{John04} If we approximate this background spin order as an SDW order, the situation is represented as a system with an SDW order with merons. Furthermore, in the pseudogap phase, vortex-like excitations and enhanced diamagnetism have been observed;\cite{ong} they may be explain by currents around merons we will discuss in the present work.

Now let us consider the ground state vector for an SDW ordered state given by
\begin{eqnarray}
\!|\Phi_{\rm SDW}(\xi) \rangle \!=\!\prod_{{\bf k}} (\cos \theta_{{\bf k}} e^{\!-\!i\xi/2}c_{{\bf k} \uparrow}^{\dagger}
\!+\!\sin \theta_{{\bf k}} e^{i\xi/2}c_{{\bf k}\!+\!{\bf Q} \downarrow}^{\dagger}\!)\! |{\rm vac} \rangle,\! \! \! \!
\label{sdw1}
\end{eqnarray}
where $c_{{\bf k} \sigma}^{\dagger}$ is a creation operator for a state with wave vector ${\bf k}$ and spin $\sigma$, 
${\bf Q}$ is the Fermi surface nesting vector, and $\xi$ is the phase of the SDW order parameter.

A salient feature of this state vector is that two states with different $\xi$ values are physically distinct unless the difference is a multiple of $2\pi$. Thus, the ground state is degenerate and labeled by $\xi$. 
This degeneracy gives rise to a possibility to have states with a spatially and temporally varying $\xi$.

Using a fermion field operator
\begin{eqnarray}
\hat{\psi}_{\sigma}({\bf r})=\sum_{{\bf k}, \sigma} c_{{\bf k}\sigma}\phi_{{\bf k} \sigma}({\bf r}),
\label{f1}
\end{eqnarray}
where basis functions are given by $\phi_{{\bf k} \sigma}({\bf r})=V^{-1/2}e^{i{\bf k}\cdot {\bf r}}$ ($V$ is the volume of the system),
the spin density for $ |{\Phi}_{\rm SDW} \rangle $ is calculated as
\begin{eqnarray}
s_x({\bf r})&=&\langle{\Phi}_{\rm SDW}|{ 1 \over 2}\left(\hat{\psi}^{\dagger}_{\uparrow}({\bf r})\hat{\psi}_{\downarrow} ({\bf r})
+\hat{\psi}^{\dagger}_{\downarrow}({\bf r})\hat{\psi}_{\uparrow}({\bf r})\right) |{\Phi}_{\rm SDW} \rangle
\nonumber
\\
&=&|{\bf s}| \cos({\bf Q}\cdot{\bf r}+\xi),
\\
s_y({\bf r})&=&\langle{\Phi}_{\rm SDW}|{ 1 \over {2i}}\left(\hat{\psi}^{\dagger}_{\uparrow}({\bf r})\hat{\psi}_{\downarrow}({\bf r}) 
-\hat{\psi}^{\dagger}_{\downarrow}({\bf r})\hat{\psi}_{\uparrow}({\bf r})\right) |{\Phi}_{\rm SDW} \rangle 
\nonumber
\\
&=&|{\bf s}| \sin({\bf Q}\cdot{\bf r}+\xi),
\\
s_z({\bf r})&=&\langle{\Phi}_{\rm SDW}|{ 1\over 2}\left(\hat{\psi}^{\dagger}_{\uparrow} ({\bf r})\hat{\psi}_{\uparrow} ({\bf r})
-\hat{\psi}^{\dagger}_{\downarrow} ({\bf r})\hat{\psi}_{\downarrow}({\bf r})\right) |{\Phi}_{\rm SDW} \rangle 
\nonumber
\\
&=&0,
\end{eqnarray}
with $|{\bf s}|$ given by
\begin{eqnarray}
|{\bf s}|=V^{-1}\sum_{\bf k} \cos \theta_{\bf k} \sin \theta_{\bf k}.
\end{eqnarray}

For spatially modulating $\xi$, the state vector Eq.~(\ref{sdw1}) is modified as
\begin{eqnarray}
|\tilde{\Phi}_{\rm SDW} \rangle =\prod_{{\bf k}} (\cos \theta_{{\bf k}}(\hat{\bf r}) \tilde{c}_{{\bf k} \uparrow}^{\dagger}
+\sin \theta_{{\bf k}}(\hat{\bf r}) \tilde{c}_{{\bf k}+{\bf Q} \downarrow}^{\dagger}) |{\rm vac} \rangle,
\label{sdw2}
\end{eqnarray}
where $\hat{\bf r}$ is the position operator. It is replaced by the argument of a field operator with which the contraction of the accompanying creation operator is taken; for example, 
$
\contraction{}{\hat{\psi}({\bf r})}{\cos \theta_{\bf k}(\hat{\bf r})}{{\tilde{c}}^{\dagger}}{}
{{\hat{\psi}}({\bf r})}{\cos \theta_{\bf k}(\hat{\bf r})}{{\tilde{c}}^{\dagger}_{{\bf k} \uparrow}}
=\contraction{}{\hat{\psi}({\bf r})}{\cos \theta_{\bf k}(\hat{\bf r})}{{\tilde{c}}^{\dagger}}{}{\hat{\psi}}({\bf r}) \cos \theta_{\bf k}({\bf r}){\tilde{c}}^{\dagger}_{{\bf k} \uparrow}.
$

New fermion operators $\tilde{c}_{{\bf k} \uparrow}^{\dagger}$ and $\tilde{c}_{{\bf k} \downarrow}^{\dagger}$ are
creation operators for basis functions that take into account the spatial variation of $\xi$,
\begin{eqnarray}
\tilde{\phi}_{{\bf k} \uparrow}({\bf r})=e^{i {1 \over 2}(\chi({\bf r})-\xi({\bf r})) }\phi_{{\bf k} \uparrow}({\bf r})
\end{eqnarray}
 and 
\begin{eqnarray}
\tilde{\phi}_{{\bf k} \downarrow}({\bf r})=e^{i {1 \over 2}(\chi({\bf r})+\xi({\bf r})) }\phi_{{\bf k} \downarrow}({\bf r}),
\end{eqnarray}
respectively; and the field operator is now given by
\begin{eqnarray}
\hat{\tilde{\psi}}_{\sigma}({\bf r})=\sum_{{\bf k}, \sigma} \tilde{\phi}_{{\bf k} \sigma}({\bf r})\tilde{c}_{{\bf k} \sigma},
\label{f2}
\end{eqnarray}
where the $U(1)$ phase factor $e^{i \chi({\bf r})/2}$ that does not alter the spin density is added. Without it, the new basis functions are not single-valued: because the value of the phase $\xi({\bf r})$ changes by $2\pi$ after a circular transport around a meron, the sign change of $e^{i\xi({\bf r})/2}$ and $e^{-i\xi({\bf r})/2}$ occurs; thus, $e^{i \chi({\bf r})/2}$ is needed to compensate this sign change. The compensation may be done, for example, by choosing $\chi$ to be $\chi({\bf r})=\xi({\bf r})$, but the real $\chi$ solution should be determined by the minimum total energy condition. Note that if $\chi_1$ is a solution, $-\chi_1$ is also a solution by the time-reversal invariance of the Hamiltonian.

The appearance of the phase factor $e^{i \chi({\bf r})/2}$ may also be understood in the following way: instead of Eq.~(\ref{sdw2}), the straightforward inclusion of the spatially varying $\xi$ phase yields
\begin{eqnarray}
|\bar{\Phi}_{\rm SDW} \rangle =\prod_{{\bf k}} (\cos \theta_{{\bf k}}(\hat{\bf r}) \bar{c}_{{\bf k} \uparrow}^{\dagger}
+\sin \theta_{{\bf k}}(\hat{\bf r}) \bar{c}_{{\bf k}+{\bf Q} \downarrow}^{\dagger}) |{\rm vac} \rangle,
\end{eqnarray}
with fermion operators $\bar{c}_{{\bf k} \uparrow}^{\dagger}$ and $\bar{c}_{{\bf k} \downarrow}^{\dagger}$ for 
\begin{eqnarray}
\bar{\phi}_{{\bf k} \uparrow}({\bf r})=e^{-i {1 \over 2}\xi({\bf r})) }\phi_{{\bf k} \uparrow}({\bf r})
\label{phi1b}
\end{eqnarray}
and
\begin{eqnarray}
\bar{\phi}_{{\bf k} \downarrow}({\bf r})=e^{i {1 \over 2}\xi({\bf r}) }\phi_{{\bf k} \downarrow}({\bf r}),
\label{phi2b}
\end{eqnarray}
respectively.
Then, the wave function
\begin{eqnarray}
\bar{\Psi}({\bf r}_1, \cdots, {\bf r}_N; \xi({\bf r}))=\langle {\bf r}_1, \cdots, {\bf r}_N|\bar{\Phi}_{\rm SDW} \rangle
\end{eqnarray}
is a multiple-valued one due to the multi-valuedness of $e^{i \xi({\bf r})/2}$.

The total wave function, a product of a wave function for the above internal state and a wave function for a motion of a system as a whole, is given by
\begin{eqnarray}
e^{i \sum_{j} \bar{\chi}({\bf r}_j)} \bar{\Psi}({\bf r}_1, \cdots, {\bf r}_N; \xi({\bf r})),
\end{eqnarray}
where $e^{i \sum_{j} \bar{\chi}({\bf r}_j)}$ is the wave function for the whole system motion. The total wave function must be single-valued; thus, $\bar{\chi}$ must have the multi-valuedness that compensates the muti-valuedness of $\bar{\Psi}$. 

If we choose $\bar{\chi}$ to be ${ 1 \over 2}\chi$, the single-valued total wave function is obtained; and the resulting wave function is,
\begin{eqnarray}
e^{i\sum_{j} { 1 \over 2} \chi({\bf r}_j)} \bar{\Psi}({\bf r}_1, \cdots, {\bf r}_N; \xi({\bf r}))
=\langle {\bf r}_1, \cdots, {\bf r}_N|\tilde{\Phi}_{\rm SDW} \rangle.
\end{eqnarray}
Therefore, $e^{i\sum_{j} { 1 \over 2} \chi({\bf r}_j)}$ can be regarded as a wave function for the motion of the system as a whole.

The nontrivial phase factor $e^{i \chi({\bf r})/2}$ gives rise to a circular current around the meron. 
We will call it, a ``meron current". Corresponding to the two solutions $\chi_1$ and $-\chi_1$, there are two choices in the direction of the current.

In the following we assume that the spatial variation of $\theta_{\bf k}$'s are negligibly small.

The electric current density for $ |\tilde{\Phi}_{\rm SDW} \rangle$ is calculated as
\begin{eqnarray}
{\bf j}({\bf r})&=&-i\sum_{\sigma}{ {q\hbar} \over m }\langle\tilde{\Phi}_{\rm SDW}|\hat{\tilde{\psi}}^{\dagger}_{\sigma}({\bf r})\nabla \hat{\tilde{\psi}}_{\sigma} ({\bf r})|\tilde{\Phi}_{\rm SDW} \rangle 
\nonumber
\\
&=&{{\hbar q \rho} \over {2m}} \nabla \chi,
\label{current}
\end{eqnarray}
where $m$ is an effective mass, $\rho$ is the particle density, and the current from the ordinary SDW state without merons is calculated to be zero. 

Let us examine a current generation by a collection of merons,
\begin{eqnarray}
\xi({\bf r})=\sum_c \tan^{-1}{{y-y_c} \over {x-x_c}},
\end{eqnarray}
which gives rise to spin density modulation,
\begin{eqnarray}
s_x({\bf r})&=&\langle\tilde{\Phi}_{\rm SDW}|{ 1 \over 2}\left(\hat{\tilde{\psi}}^{\dagger}_{\uparrow}({\bf r})\hat{\tilde{\psi}}_{\downarrow} ({\bf r})
+\hat{\tilde{\psi}}^{\dagger}_{\downarrow}({\bf r})\hat{\tilde{\psi}}_{\uparrow}({\bf r})\right) |\tilde{\Phi}_{\rm SDW} \rangle 
\nonumber
\\
&=&|{\bf s}| \cos({\bf Q}\cdot{\bf r}+\xi({\bf r})),
\\
s_y({\bf r})&=&\langle\tilde{\Phi}_{\rm SDW}|{ 1 \over {2i}}\left(\hat{\tilde{\psi}}^{\dagger}_{\uparrow}({\bf r})\hat{\tilde{\psi}}_{\downarrow}({\bf r}) 
-\hat{\tilde{\psi}}^{\dagger}_{\downarrow}({\bf r})\hat{\tilde{\psi}}_{\uparrow}({\bf r})\right) |\tilde{\Phi}_{\rm SDW} \rangle 
\nonumber
\\
&=&|{\bf s}| \sin({\bf Q}\cdot{\bf r}+\xi({\bf r})),
\\
s_z({\bf r})&=&\langle\tilde{\Phi}_{\rm SDW}|{ 1\over 2}\left(\hat{\tilde{\psi}}^{\dagger}_{\uparrow} ({\bf r})\hat{\tilde{\psi}}_{\uparrow} ({\bf r})
-\hat{\tilde{\psi}}^{\dagger}_{\downarrow} ({\bf r})\hat{\tilde{\psi}}_{\downarrow}({\bf r})\right) |\tilde{\Phi}_{\rm SDW} \rangle 
\nonumber
\\
&=&0.
\end{eqnarray}

\begin{figure}
\begin{center}
\includegraphics[scale=0.5]{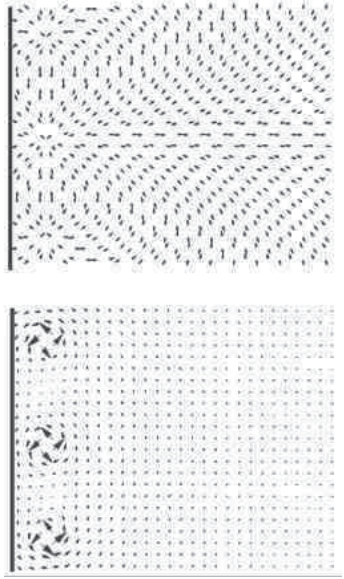}
\end{center}
\caption{\label{fig:meron} Top: a spin texture generated by three merons and an SDW order with ${\bf Q}=(\pi,\pi)$ in the two-dimensional space. The left solid line indicates the left edge of the system.
Bottom: a current induced by merons, where $\chi$ is chosen to be $\chi=\xi$.}
\end{figure}

An example of a two dimensional spin texture generated by an SDW order with three merons is depicted in Fig.~\ref{fig:meron}. A corresponding current is also shown in Fig.~\ref{fig:meron} by taking $\chi=\xi$.

If the balance of upward and downward currents is destroyed, a collection of meron currents may give rise to a macroscopic current.
To investigate such a possibility, we first examine the meaning of the nontrivial phase $\chi$. 

An appearance of the nontrivial phase actually means an appearance of a magnetic field. The phase factor $e^{i { 1 \over 2} \chi({\bf r})}$ changes the derivative to a {\em covariant derivative},
\begin{eqnarray}
\nabla f({\bf r}) \rightarrow \left(\nabla+i { 1 \over 2} \nabla \chi \right) f({\bf r}),
\label{cdev}
\end{eqnarray}
which may be compared with the covariant derivative in the presence of a magnetic field
${\bf B}=\nabla \times {\bf A}$,
\begin{eqnarray}
 \left( \nabla-i { q \over {\hbar c}} {\bf A}\right) f({\bf r}).
\label{cdev2}
\end{eqnarray}

If we define a fictitious vector potential ${\bf A}_{\rm fic}$ by
\begin{eqnarray}
{\bf A}_{\rm fic}=-{{\hbar c} \over {2q}}\nabla \chi,
\label{fic}
\end{eqnarray}
the correspondence between the r.h.s of Eq.~(\ref{cdev}) and Eq.~(\ref{cdev2}) suggests
that, ${\bf A}_{\rm fic}$ plays a role of the vector potential for a magnetic field.

It is known that the effect of a magnetic field on wave functions amounts to assign a nonintegrable phase factor to every loop along which the charged particle travels in semiclassical sense.\cite{Yang75}  The information about this is encoded in ${\bf A}$ as a {\em connection}. The vector potential ${\bf A}_{\rm fic}$ does the same assignment for electrons in the SDW band.

In other words, for the SDW electronic state with merons that give rise to the phase $\xi$, the wave function is the one augmented by a nonintegrable phase factor that is exactly the same one produced by the magnetic field given by $\nabla \times {\bf A}_{\rm fic}$. Thus, as to the effect on electrons in the SDW band, ${\bf A}_{\rm fic}$ plays the same role as a true vector potential, and $\nabla \times {\bf A}_{\rm fic}$ as a true magnetic field. 

Therefore, we have
\begin{eqnarray}
{\bf B}=\nabla \times{\bf A}_{\rm fic}=-{{\hbar c} \over {2q}}\nabla \times \nabla \chi,
\label{B}
\end{eqnarray}
which is not zero due to the multi-valuedness of $\chi$. It is also worth mentioning that
if ${\bf A}_{\rm fic}$ is given by $-{{\hbar c} \over {q}}\nabla \chi$ instead of $-{{\hbar c} \over {2q}}\nabla \chi$, the phase factor for every loop is $+1$ (the phase is $2\pi n$ where $n$ is an integer); thus, a nonintegrable phase factor does not arise.

Using ${\bf A}_{\rm fic}$, Eq.~(\ref{current}) is written as
\begin{eqnarray}
{\bf j}({\bf r})=-{{q^2 \rho} \over {mc}} {\bf A}_{\rm fic}.
\label{c2}
\end{eqnarray}
This is the London equation\cite{London} with the real vector potential replaced by the fictitious one. 

From Eqs.~(\ref{B}) and (\ref{c2}), and one of Maxwell's equations 
\begin{eqnarray}
\nabla \times \overline{\bf B}={{4 \pi} \over c}\overline{\bf j},
\label{max}
\end{eqnarray}
we obtain an equation derived by London:\cite{London} 
\begin{eqnarray}
\nabla^2\overline{\bf B}({\bf r})={ 1 \over {\lambda_L^2} }\overline{\bf B},
\label{expell}
\end{eqnarray}
where $\overline{\bf A}_{\rm fic}({\bf r})$, $\overline{\bf B}({\bf r})$ and $\overline{\bf j}({\bf r})$ denote averages of 
${\bf A}_{\rm fic}({\bf r})$, ${\bf B}({\bf r})$ and ${\bf j}({\bf r})$ over a small region around ${\bf r}$ (we will denote the similar average value for a quantity $Q$ as $\overline{Q}$),
and $\lambda_L$ is the London penetration depth\cite{London}  given by
\begin{eqnarray}
\lambda_L=({ {mc^2} \over {4 \pi q^2 \rho}  })^{1/2}.
\end{eqnarray}
As is well-known Eq.~(\ref{expell}) describes an expulsion of a magnetic field.

An equation that relates the average magnetic field to the average meron current is given from Eq.~(\ref{B}) as
\begin{eqnarray}
\overline{\bf B}={{\Phi_0} \over {2\pi}}\overline{\nabla \times \nabla \chi},
\label{chi1}
\end{eqnarray}
where $\Phi_0$ is a fluxoid given by
\begin{eqnarray}
\Phi_0={{h c} \over {2e}}.
\end{eqnarray}
Here, electron charge $-e=q$ is used.

Equation (\ref{chi1}) has a simple meaning: if ${ 1 \over {2\pi}} \nabla \times \nabla \chi$ is integrated over a surface $S$ with a boundary $C$, we have
\begin{eqnarray}
{ 1 \over {2\pi}}\int_S\nabla \times \nabla \chi \cdot d{\bf S}={ 1 \over {2\pi}}\oint_C \nabla \chi \cdot d{\bf r}={ 1 \over {2\pi}}\oint_C d \chi,
\label{SMCW}
\end{eqnarray}
where the last expression is equal to the difference of the number of counterclockwise meron current centers (each center contributes $+1$) and that of clockwise ones (each center contributes $-1$) within the loop $C$. We call the quantity in Eq.~(\ref{SMCW}), `` the sum of meron current winding numbers (SMCW)".
Therefore, Eq.~(\ref{chi1}) is actually $\Phi_0$ times an average of the SMCW in units area normal to the magnetic field.

Equations~(\ref{max}), (\ref{expell}), and (\ref{chi1}) comprise a system of equations for a stable macroscopic current. Note that when the three equations are consistently satisfied, $\overline{\bf B}$ is considered to be a true magnetic field produced by $\overline{\bf j}$ through Eq.~(\ref{max}).

\begin{figure}
\begin{center}
\includegraphics[scale=0.5]{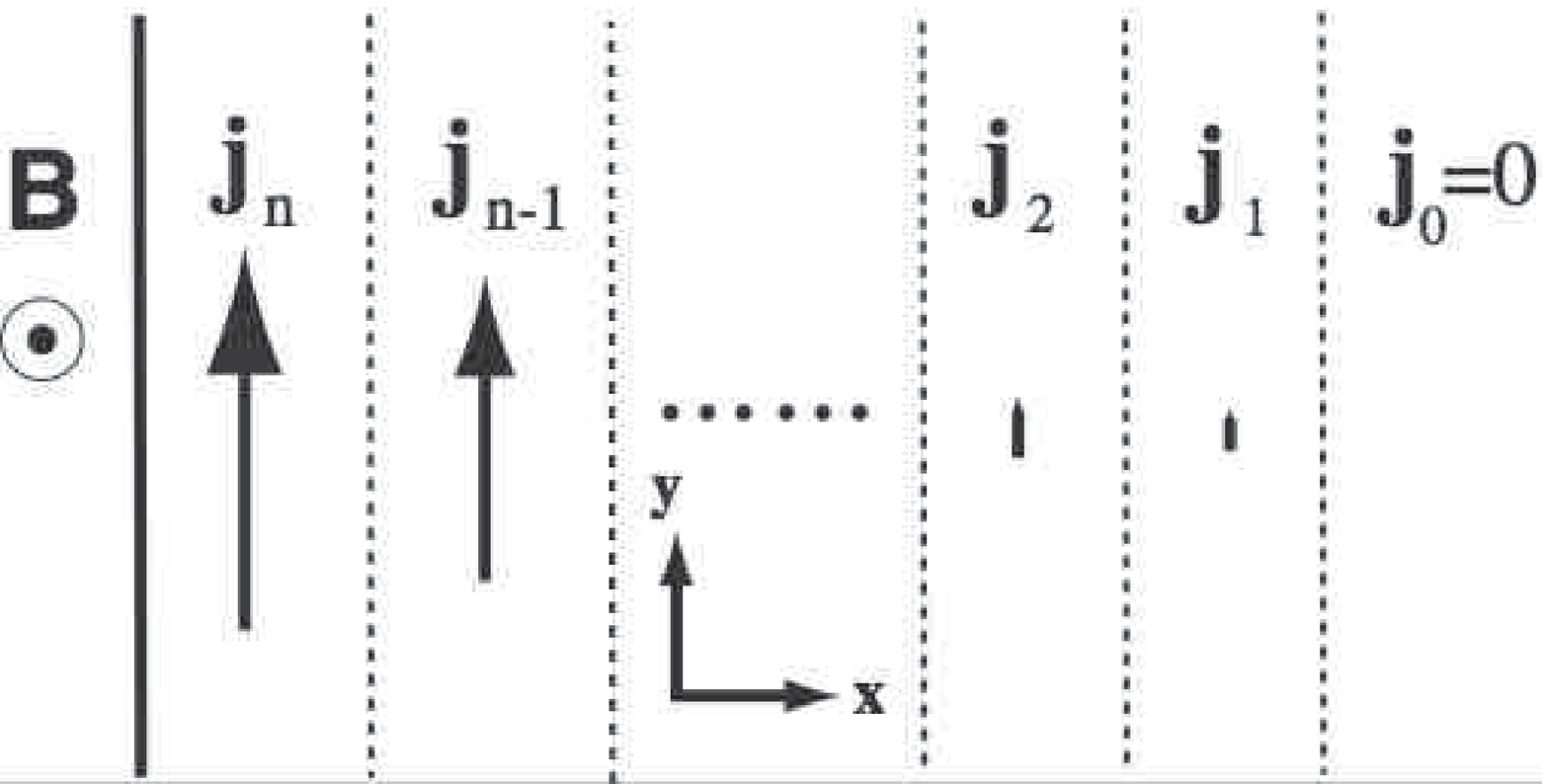}
\end{center}
\caption{\label{fig:Line} A current distribution generated by sheets of meron current centers. The system is uniform in the $z$ direction. Each sheet of meron current centers is represented by a dotted line. It causes a jump in the current. The current is a diamagnetic one, which decays to zero as the distance from the surface increases.}
\end{figure}

Let us obtain a solution for a semi infinite system that extends in the $x > 0$ region (Fig.~\ref{fig:Line}). A magnetic field points in the $z$ direction. 
The average magnetic field is calculated from Eq.~(\ref{expell}) as
\begin{eqnarray}
\overline{B}_z(x)=\overline{B}_z(0)e^{-{ x \over {\lambda_L}}},
\label{s1}
\end{eqnarray}
which decays as $x$ increases.  

From Eq.~(\ref{chi1}), an average of the SMCW
 is obtained as
\begin{eqnarray}
\rho_{\rm m}(x)={{\overline{B}_z(0)} \over {\Phi_0}}e^{-{ x \over {\lambda_L}}}.
\label{s2}
\end{eqnarray}

According to Eq.~(\ref{max}), the diamagnetic current flows in the $y$ direction, and its relation to $\rho_m$ is obtained as
\begin{eqnarray}
\overline{j}_y(x)={{\Phi_0 c} \over {4 \pi \lambda_L}}\rho_{\rm m}(x).
\end{eqnarray}

The diamagnetic current may be considered to be generated by sheets of meron current centers: from Eqs.~(\ref{fic}) and (\ref{c2}), we have
\begin{eqnarray}
\nabla \times \overline{\bf j}=-{ {\Phi_0 c} \over {8 \pi^2 \lambda_L^2} } \overline {\nabla \times \nabla \chi},
\end{eqnarray}
which is analogous to an equation in hydrodynamics given by
\begin{eqnarray}
\nabla \times {\bf u}=\boldsymbol{\omega},
\label{hydo}
\end{eqnarray}
where ${\bf u}$ and $\boldsymbol{\omega}$ are velocity and vorticity in a fluid, respectively.
We see correspondence between $\overline{\bf j}$ and ${\bf u}$, and $\overline{\nabla \times \nabla \chi}$ and $\boldsymbol{\omega}$. 
It is known in hydrodynamics that a vortex sheet causes a jump in the fluid velocity,\cite{vortex} which in the present case means that a sheet of meron current centers causes a jump in the current density. 

The current should be zero deep inside the sample; a small current that would flow if the electric resistivity of the sample is zero does not arise in real materials.
Starting from there, a successive positive current jumps through sheets of meron current centers toward the surface leads to a current density distribution that decays as $x$ increases (Fig.~\ref{fig:Line}). Actually, the current density jump is an extreme ideal case; in reality it varies gradually and continuously.

So far we have been considering the case with a collection of merons only; however, the formation of meron-antimeron pairs is energetically more favorable since the background SDW order is not destroyed in the region away from them.\cite{John04} 
Each meron current around a meron or an antimeron has two choices in its directions. When a macroscopic current is present, one of them is chosen.
Merons or antimerons are not easily destroyed at sufficiently low temperature because they are topological objects, thus the destruction of them requires a cooperative movement of a large number of electrons.
Major dissipative processes for the diamagnetic current will be single particle excitations and meron flow.
When these processes are negligible, the diamagnetic current should persist.

In summary, we have investigated a system of an SDW order with merons. It is shown that a stable circular current appears around a meron. A collection of such currents will gives rise to a diamagnetic current. If the meron density is sufficiently large, a perfect diamagnetism is realized. The stability of the diamagnetic current is attributed to the topological nature of merons, and if dissipative processes such as 
single particle excitations and meron flow are negligible, the current will be a persistent one.

\section*{Acknowledgment}
The author acknowledges helpful comments from S. Sugano and Y. Takada. He is also grateful to A. Fujimori for useful information.

\end{document}